\documentclass[smallcondensed]{svjour3_mh}     
\smartqed  

\usepackage{fix-cm}

\usepackage{amssymb,amsmath}

\usepackage{graphicx}

\newcommand{\R}{\mathbb{R}}
\newcommand{\grad}{\nabla}
\newcommand{\oneover}[1]{\frac{1}{#1}}
\newcommand{\eps}{\epsilon}
\newcommand{\bb}[1]{\mathbf{#1}}
\newcommand{\e}[1]{^{(#1)}}

 \journalname{arxiv}

\begin{document}

\title{Sticky-sphere clusters}

\titlerunning{Sticky-sphere clusters}

\author{Miranda Holmes-Cerfon}
\institute{M. Holmes-Cerfon \at
              \email{holmes@cims.nyu.edu} 
}

\date{}

\maketitle

\begin{abstract}
Nano- and microscale particles, such as colloids, commonly interact over ranges much shorter than their diameters, so it is natural to treat them as ``sticky,'' interacting only when they touch exactly. 
The lowest-energy states, free energies, and dynamics of a collection of $n$ particles can be calculated in the sticky limit of a  deep, narrow interaction potential. 
This article surveys the theory of the sticky limit, 
explains the correspondence between theory and experiments on colloidal clusters, and outlines areas where the sticky limit may bring new insight. 
\end{abstract}


\section{Introduction}

What can a small cluster of particles say about the materials we encounter in everyday life? While we cannot 
bang it with a hammer, wrap it around our shoulders, or throw it a ball, 
the information contained in the ground states of small systems is nevertheless critical to explaining many physical and biological properties of larger ones. 
Condensed-matter phenomena such as nucleation, 
the glass transition, 
gelation, 
epitaxial growth, 
aging, 
and the structure of liquids 
all have explanations rooted in the geometrically possible ways to arrange a small collection of particles 
\cite{Frank:1952eg,Stillinger:1984tr,Nelson:1989ij,Doye:1996vn,Sedgwick:2004di,PatrickRoyall:2008fz,Yunker:2009jd,Ganapathy:2010be}. 
These possibilities also act as constraints on biological systems like proteins, viruses, chromatin, and microtubules,  
that fold, self-assemble, metabolize, or self-replicate.  
Small clusters have been used to design synthetic systems that perform these functions, bringing insight into the geometrical origins of biological complexity \cite{Wang:2011gv,Zeravcic:2014ev}.  
Such synthetic systems are also of independent interest as we seek to design materials with new properties that may assemble or heal themselves \cite{Manoharan:2004jk,Fan:2010jn,Hormoz:2011ir,Schade:2013ee,Zeravcic:2014it}.

For many of the phenomena above it is natural to considers particles that interact over distances much smaller than the diameter of the particles. Such short-ranged interactions occur for a wide range of nano- and microscale particles, like colloids, 
where longer-ranged interactions such as electrostatic forces may be screened by ions in the fluid medium \cite{Lu:2013dn,Manoharan:2015ko}. 
Common methods to create short-ranged attractive interactions include adding a depletant to a solution \cite{Asakura:1954jy} or coating the particles with strands of complementary DNA, which acts like velcro when they get close enough \cite{Nykypanchuk:2008cp,Dreyfus:2009gl,Rogers:2011et,Macfarlane:2011fh,DiMichele:2013bw}. 
Colloids are convenient systems with which to study material behaviour because while they can be small enough to be thermally excited, and they can be buoyancy matched to be suspended in a fluid, they are still big -- big enough that they can be treated theoretically as classical bodies, and big enough that they can be studied experimentally more easily than atoms or molecules \cite{Lu:2013dn}.  
There is also an exciting possibility of using colloids to design new materials, since they can be synthesized to have a plethora of shapes, sizes and interaction structures so the parameter space of building blocks is very large \cite{Sacanna:2013ge}.

This review describes the recent progress in understanding small clusters of particles interacting with a short-ranged attractive potential, focusing primarily on modeling clusters of colloids. 
 It describes a theoretical  framework, the computational apparatus that supports it, 
 and experimental measurements that validate this framework. It does little to explain how this framework may be  applied to glean insight into scientific questions, and it  does not broach the significant literature on simulating systems that are close to sticky,  e.g. \cite{MartinezVeracoechea:2011eoa,Romano:2012bz,Li:2013he,Millan:2014de}.
One reason for this focus is that the framework is relatively new and under development. Another is that the ideas and tools are expected to apply to more general systems than clusters, like jammed or glassy systems \cite{OHern:2002bsa,Boolchand:2005bi}, silicates \cite{Hammonds:1996wy}, or origami \cite{Pandey:2011jj,Demaine:2007jh,Silverberg:2014dn}, which can be modeled as objects linked by soft, stiff constraints, even when the interactions are purely repulsive. 
It is hoped that by focusing on the theoretical apparatus, 
 connections to other fields may be easier to make.

The framework to be described is different from the traditional approach to energy landscapes, which, in its simplest form, characterizes a high-dimensional energy landscape by a set of local minima and transition states \cite{Stillinger:1984tr,Wales:2012dd}. 
The local minima represent metastable states where a system spends long amounts of time, and the height of the transition states (usually saddle points) determines the rate of transition between minima through the Arrhenius formula. There are many sophisticated techniques for computing the local minima and transition states and for building upon these ideas, which together have yielded an extremely powerful set of methods that have brought insight to a great many atomic, molecular, and condensed-matter systems (e.g. \cite{Wales:2003}, and references therein.)

Yet, for colloidal clusters these methods suffer from a few disadvantages.  
One is that the energy landscape depends sensitively on the interaction potential, 
which is often not well known in soft matter systems. Even when it can be estimated, the computations must be re-done for each distinct potential. 
In addition, because the methods are based on searching the landscape 
stochastically, there is no way to guarantee they have found all the important pieces of it. 
Finally, describing the dynamics by the heights and locations of the saddle points, or even the full transition paths themselves, becomes less accurate as the potential narrows, except when the temperature is unrealistically low. 

\begin{figure}
\begin{center}
\includegraphics[width=0.8\linewidth]{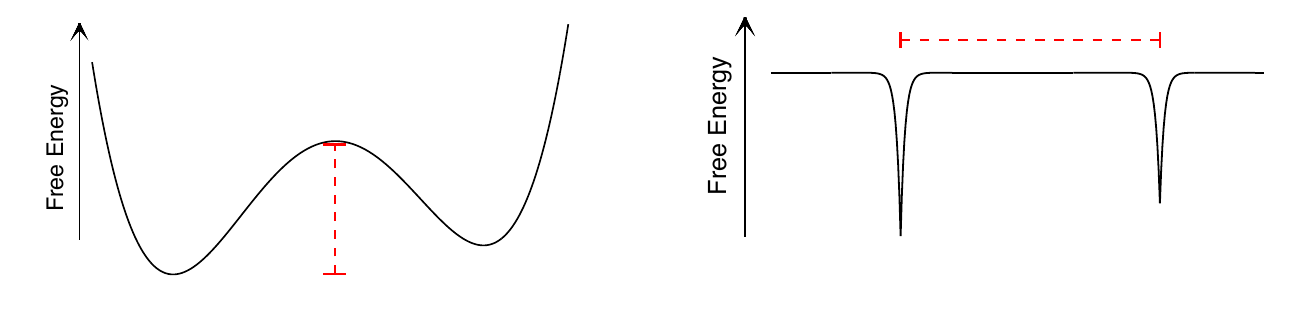}
\end{center}
\caption{Left: traditional schematic of an energy landscape. Transitions between local minima occur at low-lying saddle points, whose height determines the rate of transition. Right: colloidal energy landscape. The regions between the local minima are nearly flat so the dynamics on these regions are mostly diffusive. Therefore, the size and shape of each region are also important factors in determining the transition rate. Adapted with permission from \cite{HolmesCerfon:2013jw}. }\label{fig:landscape}
\end{figure}

The reason this dynamical description breaks down is illustrated in Figure \ref{fig:landscape}. On the left is a traditional schematic of an energy landscape, usually drawn as a hilly surface. The local minima live in smooth basins of attraction, which are joined together by smooth saddle points, whose heights determine the rate of transition between the basins. On the right is a schematic of a colloidal energy landscape, where the interactions between particles are short-ranged. The basins of attraction are much narrower, and the regions in between are much flatter in comparison. No amount of information at a single point (height, curvature, location, etc) will determine the rate of transition between the basins; one needs to know something about the size and shape of the whole transition region. 

This article considers the energy landscape and dynamics of a collection of spherical particles when the range of interaction goes to zero -- the so-called ``sticky limit.'' 
In this limit the free energy landscape is given by a set of geometrical manifolds (shapes of different dimensions), plus a single parameter that incorporates all system-dependent information such as the interaction potential and temperature. 
The manifolds depend only on the geometry of the particles,  and combined with the dynamical equations defined on the manifolds, 
provide the starting point from which any quantity characterizing the system -- equilibrium or nonequilibrium -- can be computed for arbitrary potentials. 

We proceed as follows. In section \ref{sec:enum} we describe the set of rigid clusters of $n$ spherical particles, which are local minima on the energy landscape in the sticky limit. In section \ref{sec:thermo}  we consider the free energy of clusters, both rigid and floppy. We summarize the theoretical predictions, show they generally agree with experimental measurements, and explain situations they cannot yet describe because of singularities in the sticky limit. Section \ref{sec:kinetics} introduces the equations describing a cluster's dynamics in the sticky limit, and shows they can be used to predict experimental transition rates. Finally, in section \ref{sec:outlook} we explain how the sticky limit  may give insight into systems other than clusters. 



\section{Rigid clusters}\label{sec:enum}

When the interaction potential between particles is extremely narrow, then to a first-order approximation it can be treated as a delta-function. 
In this case particles only interact when they are exactly touching, and the only energy barriers correspond to breaking a contact. 
If the particles are spheres with identical interactions, then there is a nice relation between clusters that are energetic local minima and the mechanical properties of a cluster as a framework. 
That is, a cluster is typically a local minimum if it has no internal degrees of freedom: it cannot move around while maintaining all contacts, except by rigid body motions. 
This means it is rigid when thought of as a framework (graph), where the sphere centers are the vertices (hinges)  and each contact is a bar (edge). 
If a cluster is floppy, i.e. not rigid, then it can typically deform until two spheres come into contact, which lowers its potential energy without crossing an energy barrier.\footnote{Another possibility is the cluster could deform continuously without forming a contact, like moving on a circle. A floppy cluster with this property would still be classified as a local minimum, albeit a degenerate one.}

Finding the local minima on this delta-function energy landscape, is therefore equivalent to finding the ways that $n$ spheres can be arranged into a rigid cluster,  a problem first suggested by  Arkus et al \cite{Arkus:2009dc}. 
This problem is conceptually more appealing than minimizing an energy function because one can  potentially prove whether the solution set is complete (e.g. \cite{Bezdek:2012if,Bezdek:2013de}.)
In this section we focus on the geometry of rigid clusters. First we define rigidity and explain how it can be efficiently tested, then we qualitatively describe the known set of rigid clusters, and finally we survey methods to find them. 

\subsection{Setup}\label{sec:rigidsetup}

Let a cluster be represented as a vector $\bb{x} = ( \bb{x}_1,  \bb{x}_2, \ldots,  \bb{x}_{n})\in \mathbb{R}^{3n}$ where $ \bb{x}_i = (x_{3i-2}, x_{3i-1}, x_{3i})$ is the center of the $i$th sphere. 
The cluster has $m$ pairs of spheres in contact $E=\{(i_1,j_1),\ldots,(i_m,j_m)\}$.
For each pair $(i,j)$ in contact there is an algebraic equation
\begin{equation}\label{eq:bonds}
| \bb{x}_i -  \bb{x}_j|^2 = d_{ij}^2, \qquad (i,j) \in E,
\end{equation}
where $d_{ij}$ is the sum of the two radii. Hereafter we consider identical spheres with unit diameters (hence $d_{ij}=1$), and additionally require that spheres not overlap so $|\bb{x}_i-\bb{x}_j|\geq 1$ for all $i\neq j$. 
This system can be represented by an adjacency matrix $A$ by setting $A_{ij} = 1$ if spheres $i,j$ are in contact, and $A_{ij} = 0$ otherwise.

A cluster is defined to be \emph{rigid} if it lies on a connected component of the solution set to \eqref{eq:bonds} that contains only rotations and translations \cite{Asimow:1978en,Connelly:1996vj,Connelly:2015dp}. 
Equivalently, a cluster is rigid if it is an isolated solution to \eqref{eq:bonds}, after factoring out rigid-body motions \cite{Demaine:2007jh}.
 Physically, being rigid means one cannot continuously deform the cluster by any finite amount while maintaining all contacts (bonds.)

\subsection{Alternative concepts of rigidity}\label{sec:rigid}

This notion of rigidity is nonlinear and there is no way to test it efficiently \cite{Demaine:2007jh}. 
In what follows we consider several alternative concepts of rigidity that are easier to test. 
These ideas are closely linked to Maxwell counting arguments used to study isostatic networks (e.g. \cite{Lubensky:2015jd}), and we will point out the correspondence. 

The first concept is  \emph{minimal rigidity},\footnote{This definition is different from that in rigidity theory, which calls a graph ``minimally rigid in dimension 3'' if it has exactly $3n-6$ edges and it has an infinitesimally rigid realization in $\R^3$ \cite{Borcea:2004dr}.}  a term introduced in this context by Arkus et al. \cite{Arkus:2009dc}.
A cluster is said to be minimally rigid if it has $3n-6$ contacts, and if each particle has at least 3 contacts. 
This comes from counting the constraints that are necessary generically to remove all degrees of freedom:  there are $3n$ variables for the sphere positions, and six rigid-body degrees of freedom, so generically one needs $3n-6$ equations to obtain an isolated solution. 
This condition is neither necessary nor sufficient for rigidity, but it is easy to test.

A richer and more rigorous concept comes from considering first- and second-order perturbations to a particular solution $\bb{x}_0$ to \eqref{eq:bonds}. 
Suppose there is a continuous path $\bb{x}(t)$ with $\bb{x}(0)=\bb{x}_0$. Taking one derivative of \eqref{eq:bonds} shows that  
\begin{equation}\label{eq:R}
R(\bb{x}_0)\bb{x}'|_{t=0} = 0,
\end{equation}
where $R(\bb{x}_0)$ is half the Jacobian of \eqref{eq:bonds}, often called the \emph{rigidity matrix}. 
A solution $\bb{x}'|_{t=0}$ is called a \emph{first-order flex} or just \emph{flex}, and the flex is  \emph{trivial} if it is an infinitesimal rigid-body motion. 
Physically, a flex is a set of velocities assigned to the particles that maintain the contacts to first order.  
Let $\mathcal{V}$ be the space of non-trivial flexes and let $\text{dim}(\mathcal{V})=N_f$. 
If $N_f=0$, then the cluster is \emph{infinitesimally rigid}, or \emph{first-order rigid}. This is sufficient for the cluster to be rigid \cite{Connelly:1996vj}.

If a cluster is not first-order rigid, then it is because either the number of contacts is too small, or the equations \eqref{eq:bonds} are linearly dependent, becoming ``tangent'' in some high-dimensional space. 
In the latter case the cluster has an interesting mechanical property: there is a set of forces one can put between the particles in contact so the cluster is in mechanical equilibrium. Such a distribution of forces is called a \emph{state of self-stress}, and can be shown to be in one-to-one correspondence with the elements in the left null space of the rigidity matrix \cite{Connelly:1996vj,Lubensky:2015jd}. Call this space $\mathcal{W}$, and let $\text{dim}(\mathcal{W}) = N_s$. 
The number of variables, contacts, flexes, and states of self-stress are related by 
the rank-nullity theorem in linear algebra as 
\begin{equation}\label{eq:maxwell}
N_f-N_s = 3n-6-m.
\end{equation}
This equation, often described as Calladine's extension \cite{Calladine:1978dy} of the Maxwell rule \cite{Maxwell:1864vl}, 
 has played an important role in the physics literature. It has been applied to a variety of materials that can be characterized by their set of contacts, such as random packings, jammed or glassy systems, or synthetic materials based on periodic frameworks (e.g. \cite{Lubensky:2015jd}, and references therein.) 
Yet, while \eqref{eq:maxwell} moves beyond minimal rigidity by characterizing additional mechanical properties, it is still a linear theory.

To move toward a nonlinear concept of rigidity we continue the Taylor expansion. 
Suppose we have a nontrivial flex $\bb{x}'|_{t=0}$, and would like to know if it extends to a finite motion. 
Taking two derivatives of  \eqref{eq:bonds} gives 
\begin{equation}\label{eq:Rp}
R(\bb{x}_0)\bb{x}''|_{t=0}= -R(\bb{x}')\bb{x}'|_{t=0}.
\end{equation}
We must solve this for $\bb{x}''|_{t=0}$. 
If we can't, then $\bb{x}'|_{t=0}$ does not extend to a second-order motion. 
If there is no nontrivial flex for which it is possible to solve \eqref{eq:Rp}, then the cluster is \emph{second-order rigid}. This is also sufficient for the cluster to be rigid \cite{Connelly:1996vj}.

Testing for second-order rigidity is too difficult, but we can strengthen the concept without losing much physics. 
Notice that, by the Fredholm alternative, we can solve for $\bb{x}''|_{t=0}$ if and only if there exists $\bb{v}\in\mathcal{V}$ such that $\bb{w}^TR(\bb{v})\bb{v} =0$ for all $\bb{w}\in\mathcal{W}$. 
The cluster is second-order rigid when this is not true: for each $\bb{v}\in\mathcal{V}$, there exists a $\bb{w}\in\mathcal{W}$ such that $\bb{w}^TR(\bb{v})\bb{v} \neq 0$. Finding a $\bb{w}$ that blocks each $\bb{v}$ separately is too hard, 
but we may be able to find 
a single $\bb{w}$ that blocks all $\bb{v}$. If there exists a $\bb{w}\in\mathcal{W}$ such that 
\begin{equation}\label{eq:pss}
\bb{w}^TR(\bb{v})\bb{v} \succ 0 \qquad \forall \;\bb{v}\in\mathcal{V},   
\end{equation}
then the cluster is clearly second-order rigid, hence rigid. A cluster which satisfies \eqref{eq:pss} is called \emph{prestress stable}.

Prestress stability is stronger than second-order rigidity, yet a large and useful step beyond linear theory. 
One major advantage is that it can be tested efficiently. Notice that the inner product in \eqref{eq:pss} can be written as $\bb{v}^T\Omega(\bb{w})\bb{v}$, where $\Omega(\bb{w})$ is a matrix constructed from the coefficients of $\bb{w}$. This matrix lives in a linear space, 
and if we restrict our attention to the set of matrices which are positive semi-definite over $\mathcal{V}$, a convex set of matrices, then our task is to find the matrix of maximal rank. 
This can be done using semidefinite programming methods (\cite{Connelly:2015dp}, and references therein.)


\subsection{The set of rigid clusters}\label{sec:clusters}

\begin{figure}
\center
\includegraphics[width=0.9\linewidth]{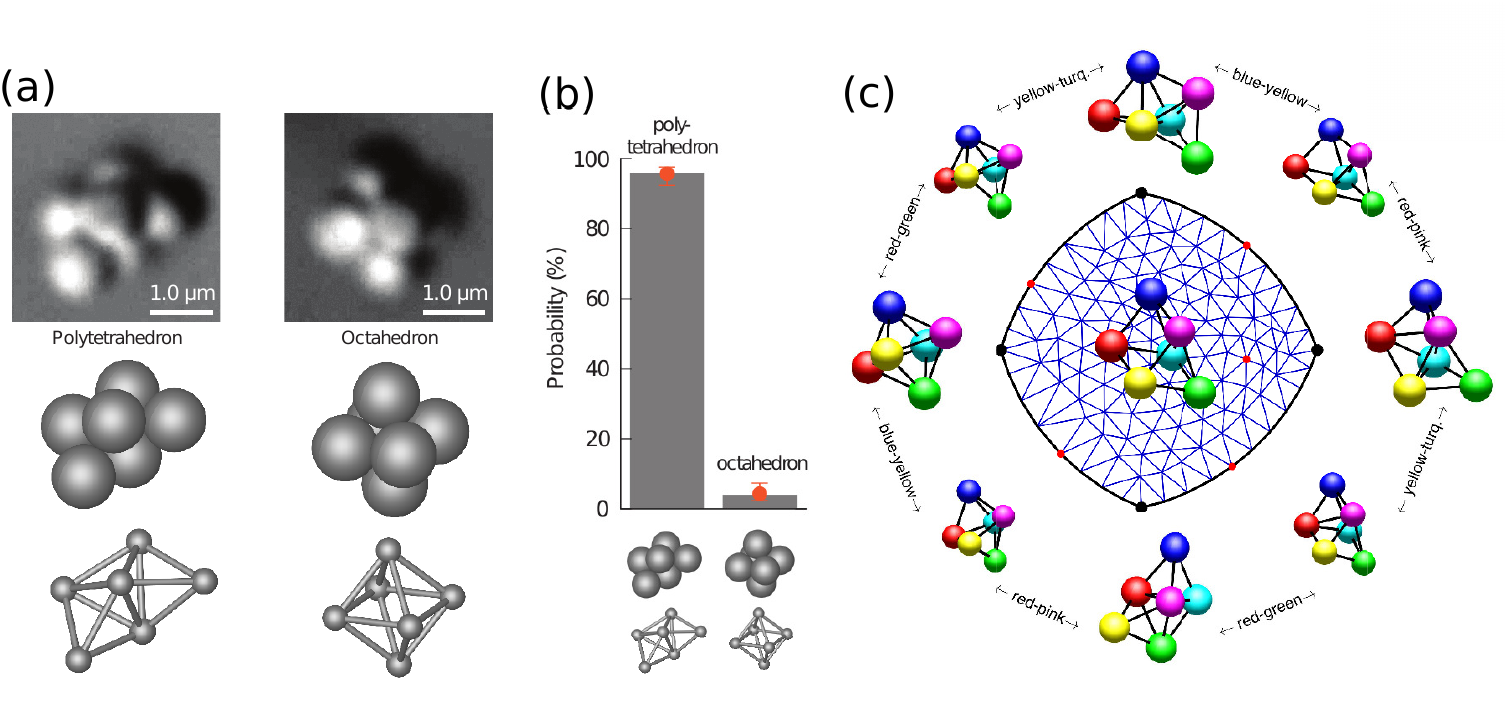}
\caption{Features of the landscape for $n=6$ spheres.  (A) Experimental images of the polytetrahedron and octahedron, and the corresponding hard-sphere packings. (B) Theoretical (bars) and experimental (red dots) equilibrium probabilities, measured in \cite{Meng:2010gsa}. (C) A two-dimensional manifold from the $n=6$ landscape. Corners are rigid clusters: one octahedron, and three polytetrahedra equivalent up to permutations. Edges are one-dimensional manifolds, formed by breaking a bond from a rigid cluster; these are the lowest-energy transition paths between rigid clusters. The interior represents all states accessible by breaking two bonds from a rigid cluster and moving on the two internal degrees of freedom. This set of states is a two-dimensional manifold that has been parameterized and triangulated in the plane.
Parts (A),(B) adapted from \cite{Meng:2010gsa}. Reprinted with permission from AAAS.
Part (C) adapted with permission from \cite{HolmesCerfon:2013jw}.}\label{fig:n6}
\end{figure}

\begin{figure}
\center
\includegraphics[width=0.9\linewidth]{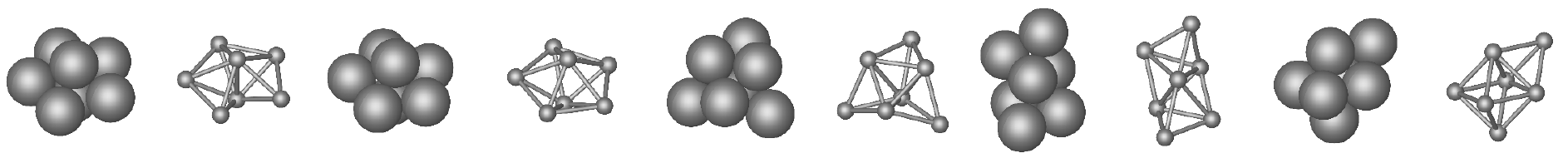}
\caption{Rigid clusters of $n=7$ spheres. The first and second clusters differ by the short path obtained by breaking the bond on the central axis of the first one. 
}\label{fig:n7}
\end{figure}

\begin{figure}
\center
\includegraphics[width=0.6\linewidth]{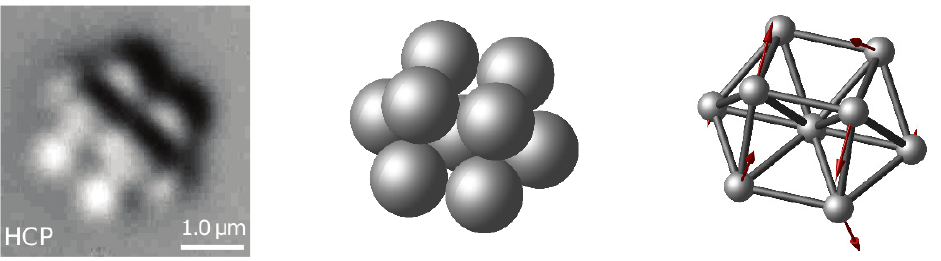}
\caption{A singular rigid cluster first occurs at $n=9$. Experimental image and corresponding sphere cluster and framework, with arrows indicating the singular motion. From \cite{Meng:2010gsa}. Reprinted with permission from AAAS.
}\label{fig:singular}
\end{figure}

We next discuss the set of known rigid clusters and some of their interesting geometrical and statistical properties. 
All clusters listed here have been tested for prestress stability \cite{HolmesCerfon:2016wa}.
The total number for each $n$ is listed in brackets, with  enantiomers lumped into a single state.

\begin{figure}
\center
\includegraphics[width=0.8\linewidth]{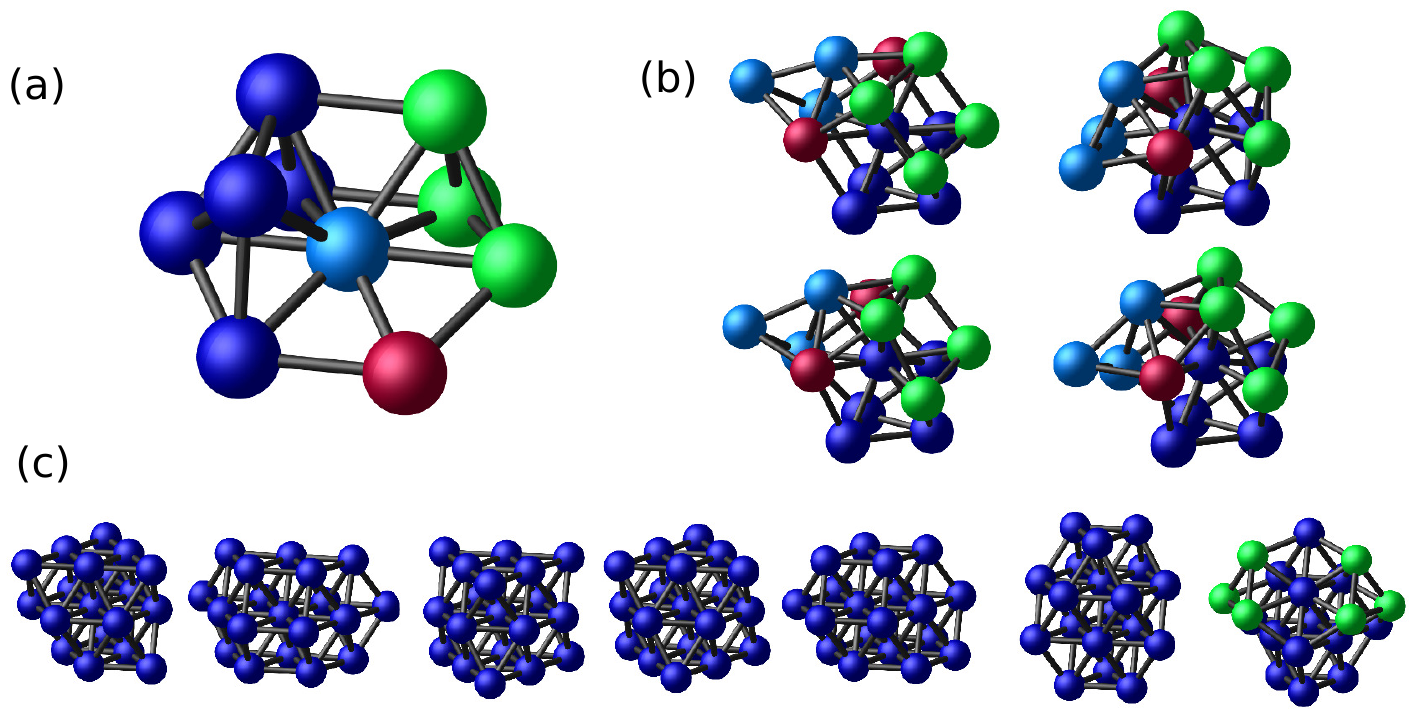}
\caption{Some interesting rigid clusters. (A) The smallest hypostatic cluster, $n=10$. (B) Four clusters with the same adjacency matrix, $n=14$, coloured to aid identification. (C) The 7 clusters for $N=19$ with the maximal number of contacts. All but one are fragments of a close-packed lattice (defects in green.) }\label{fig:clusters}
\end{figure}

\begin{itemize}
\item $\mathbf{n=3,4,5}$ (1;1;1) The sole rigid clusters are the triangle, tetrahedron, and bipyramid formed by gluing two tetrahedra together. 

\item $\mathbf{n=6}$ (2) This is the smallest interesting system because it has more than one rigid cluster: the polytetrahedron formed by gluing three tetrahedra together, and the octahedron, which does not contain any tetrahedra (Figure \ref{fig:n6}). The octahedron has 24 elements in its symmetry group while the polytetrahedron has only 2,  a fact that will be important in determining the free energy in section \ref{sec:thermo}. 

\item $\mathbf{n=7}$ (5) Three clusters are obtained by gluing a sphere to the polytetrahedron, one by gluing a sphere to the octahedron, and one cannot be decomposed into smaller rigid clusters except triangles (Figure \ref{fig:n7}.) Two clusters differ by a tiny amount. One is formed by stacking tetrahedra around a central axis, which cannot quite formed a closed loop. By breaking the contact along the central axis, the two spheres on the axis can move apart by $\approx0.05$ and the loop can close into a pentagon.  

\item $\mathbf{n=8}$ (13) All clusters but one are formed by gluing a sphere to a cluster of $n=7$. 

\item $\mathbf{n=9}$ (52) One cluster stands out because it has an infinitesimal degree of freedom. 
It is made of two bipyramids, which share a vertex and are held together by three parallel contacts (edges) (Figure \ref{fig:singular}.) When the bipyramids twist relative to each other, the lengths of the edges do not change to first order in the amount of deformation, so the twist is an infinitesimal degree of freedom. The lengths do change to second order, so the cluster is rigid. We call a cluster that is rigid but not infinitesimally rigid a \emph{singular} cluster.

\item $\mathbf{n=10}$ (263) This is the smallest system that contains both \emph{hyperstatic} clusters, those with more than the $3n-6$ contacts required generically for rigidity, as well as \emph{hypostatic} clusters, those with fewer than $3n-6$ contacts. 
There are three hyperstatic clusters and their existence is expected, since a close-packed cubic lattice 
has an average of 6 contacts per sphere.
There is one hypostatic cluster and its discovery was surprising. It is ``missing'' one contact and is shown in Figure \ref{fig:clusters}. The red sphere lies in the plane of the others it touches, a property common to many hypostatic clusters. 
A good analogy is to imagine a piece of fabric in a plane that is clamped at its boundaries; it is hard for the fabric to move perpendicular to the plane.

\item $\mathbf{n=11,12}$ (1659;11,980) The first pair of geometrically distinct clusters with the same adjacency matrix occurs at $n=11$. That this is possible is not surprising from a mathematical perspective, since a system of nonlinear equations can have multiple solutions, but it is difficult to construct examples for small $n$ by hand. 

\item $\mathbf{n=13}$ (98,529) There are now clusters with a ``caged'' sphere, with no room to make another contact. Of the 8 clusters with the maximum number of contacts, two of these contain a caged sphere: one is a fragment of a face-centered cubic (fcc) lattice, the other of a hexagonal close-packed (hcp) lattice. The latter is singular, along with one more ground state.  

\item $\mathbf{n=14}$ (895,478) This set contains a great many peculiar clusters: hypostatic clusters missing three contacts, sets of four clusters sharing the same adjacency matrix, among many others (Figure \ref{fig:clusters}).  The sheer number of clusters means the dataset can act as a catalogue to test questions about the geometrical possibilities for arranging objects into a rigid configuration, with implications beyond clusters to general graphs.
For example: is a rigid framework with more than $3n-6$ contacts always non-singular? No. Is a framework with $3n-6$ contacts always rigid? No. If a rigid framework has at least four contacts per sphere, is it always the unique solution for that adjacency matrix? No. Does an isometry of an adjacency matrix always correspond to a rotation or reflection? No. 
One can often find small examples to test geometrical conjectures.

\item $\mathbf{n=15-19}$ 
Not all clusters have been listed, but it is expected that those with the maximum number of contacts have been found. This maximum number continually increases: it is $3n+\{-1,0,1,2,3\}$ for $n=15,16,17,18,19$ respectively. There are several maximally-contacting clusters for each $n$. Almost all are fragments of a close-packed lattice, and those that aren't are usually close, with defects only on the surface (Figure \ref{fig:clusters}.) 
\end{itemize}

The total number of clusters appears to increase combinatorially with $n$ \cite{HolmesCerfon:2016wa}, as roughly $2.5(n-5)!$. 
This is faster than the exponential increase of local minima claimed for clusters with smooth potentials 
\cite{Stillinger:1984tr}, though one must be cautious in extrapolating from such small values of $n$. 
Still, the discrepancy may arise because the minimum gap between non-contacting spheres in a rigid cluster appears to become arbitrarily close to 1;  for n=14 it is $1.3\times 10^{-5}$. For a smooth potential such a small gap would cause particles rearrange to lower the overall energy, perhaps merging nearby rigid clusters into a single local minimum. 

That the lowest-energy clusters are close-packing fragments, or nearly so, is in marked contrast to clusters with a non-delta function potential such as Lennard-Jones or  atomic clusters. These are known to have special values of $n$, so-called ``magic'' numbers, where a high-symmetry icosahedral arrangement is an energetic local minimum, and this arrangement can be the lowest-energy state even for $n\approx 10^3$ \cite{Echt:1981bw,Raoult:1989dl,Wales:1997jj,Wales:2012dd}. 
Such arrangements are possible when the potential has some width because the spheres can rearrange  a little bit to create new bonds, whose additional energy more than compensates the stretching of the other bonds \cite{Doye:1996vn,Wales:2010jp}.

Interestingly, it appears that the proportion of singular clusters is nearly constant: it is 3, 2.9, 2.7, 2.5\%  for $n=11,12,13,14$ respectively \cite{HolmesCerfon:2016wa}. Whether these frequencies are significant or not in a thermal system depends on the entropy of the clusters, a question to be addressed in section \ref{sec:thermo}.

\subsection{How to find rigid clusters}

Three distinct ideas have been proposed to find the set of rigid clusters using geometrical techniques, and we now describe them. 
In addition, one may introduce a specific short-ranged potential such as the Morse potential, to observe approximate rigid clusters in simulations  \cite{Malins:2009dt} or find them by searching the energy landscape \cite{Calvo:2012bw,Morgan:2014fw}, though the range must be extremely small to find all rigid clusters \cite{Wales:2010jp}.

\subsubsection{Solving from adjacency matrices}\label{sec:arkus} 

One can imagine a brute-force method to find all rigid clusters: first, list all adjacency matrices, then, solve each system of equations for the coordinates, and finally, determine if the solution is isolated. 
This is a finite, yet Herculean task, since the number of adjacency matrices grows superexponentially with $n$ as $2^{n(n-1)/2}$. 
Yet, this is exactly what Arkus et al \cite{Arkus:2009dc,Arkus:2011tl} attempted, using an iterative method to reduce the work involved. The key step is to identify patterns in the adjacency matrix for which the distances have already been solved for analytically, or patterns that imply overlapping spheres or no solutions. A pattern that has not been seen is solved for by hand. 
Once the adjacency matrices at a given $n$ have been categorized, these become new patterns to solve or eliminate solutions for larger $n$.  For example, many clusters contain a bipyramid, so if there is a sub-matrix corresponding to the bipyramid's adjacency matrix, then these spheres have known relative positions.

Arkus et al used this approach to enumerate minimally rigid clusters of $n\leq 10$ spheres. 
Here the method reached its limits, since there were 94 patterns that had to be solved by hand. While this is potentially a rigorous, analytic way to obtain the complete list of minimally rigid clusters, the iterative step was implemented on a computer so round-off errors could cause contacts to be missed or formed extraneously. 
In addition, it is not clear whether the analytic rules were applied completely; for example, whether the authors considered the multiple possible solutions for certain patterns in the adjacency matrix. 
Their list has been corroborated by subsequent studies \cite{Hoy:2012cr,HolmesCerfon:2016wa}, with the only discrepancy being the hypostatic cluster that they did not look for.

Another group looked for minimally rigid clusters using a similar method, but instead of solving semi-analytically for each cluster, they used Newton's method with random initial conditions to find a solution of \eqref{eq:bonds}. 
They enhanced the pattern classification by drawing rules from graph theory such as the  non-embeddability of certain graphs (though not all their rules were correct \cite{Hayes:2012ty}.)  
Initially their method was applied to clusters of $n\leq 11$ spheres \cite{Hoy:2012cr}, and later with parallelization it handled $n\leq 13$ \cite{Hoy:2015hz}. 
They did not consider clusters with the same adjacency matrix, and Newton's method is not guaranteed to find as solutions, so the dataset cannot be complete. 

Methods based on adjacency matrices are limited in large part by the time it takes to list all nonisomorphic adjacency matrices initially. 
This motivates the need for a bottom-up algorithm, that builds clusters out of what is currently known, rather than starting from a larger set of possibilities and deleting. The next two methods are attempts to do this.

\subsubsection{Solving by path-following}\label{sec:mhc}

Another method to enumerate rigid clusters was based on an observation about their dynamics: typically the easiest way to get from one rigid cluster to another is to break a  contact, then deform the cluster until two spheres collide. This can be turned into an algorithm to find rigid clusters, by starting with a single rigid cluster, following all one-dimensional transition paths leading out, and repeating for all rigid clusters found at the ends. 

This algorithm was implemented numerically by Holmes-Cerfon \cite{HolmesCerfon:2016wa}, to list rigid clusters for $n\leq 14$ completely, and a subset for $n\leq 19$ which is expected to contain clusters with the maximum number of contacts. Each cluster was tested for prestress stability, so is rigid to numerical tolerance. 
Because this method tested a nonlinear notion of rigidity and did not make assumptions about the number of contacts, it found a more complete, geometrically richer set of rigid clusters. 
Of course, the method is sensitive to several numerical parameters, so it is not guaranteed to find all prestress stable clusters, nor all the one-dimensional transition paths. 
Even if it could, it would still not find all prestress stable clusters since it can only reach those connected to the starting cluster by one-dimensional paths. Indeed, Holmes-Cerfon discovered a cluster that cannot be found by this method.

A by-product of this algorithm is the set of transition paths. These have the interesting property that sometimes they are topologically circles: after a contact is broken, the cluster deforms until it forms exactly the same contact in exactly the same configuration. This suggests there could be ``circular'' floppy clusters that may deform  indefinitely without becoming rigid. These would be metastable states that should be treated as local minima, like rigid clusters. So far no method has found, or even proposed to find, a small example.

\subsubsection{Toward a complete set of rules}\label{sec:wampler}

A third idea is based on an observation by Charles Wampler  \cite{Wampler:HRGv7QI6} that 
many rigid clusters are formed by gluing together smaller ones, and one can 
derive a complete set of gluing rules to form minimally rigid clusters simply by counting degrees of freedom.  
Consider a collection of $R$ rigid clusters and $P$ isolated spheres. This has a total of $6R+3P$ degrees of freedom. Suppose we can glue together  either 
(i) two vertices (on different clusters), 
(ii)  two edges, or 
(iii) two faces. 
Additionally, we can 
(iv) add a distance constraint between two spheres on different clusters. 
If there are $V,E,F,C$ instances of each of these rules respectively, they remove a total of $3V+5E+6F+C$ degrees of freedom.
Equating the number of constraints to the number of degrees of freedom of the resulting cluster gives
\begin{equation}\label{eq:wampler}
3P+6R-C-3V-5E-6F-6=0.
\end{equation}
Each integer solution to this equation gives a different gluing rule. 
One is $\{R=1,P=1,C=3\}$, which says to glue a sphere to a rigid cluster using three contacts. This can form a large fraction of rigid clusters, and is how some of the earliest studies of energy landscapes searched for clusters \cite{Hoare:1976bb}. 
Another rule is $\{R=2,P=0,C=3,V=1\}$, which builds the $n=9$ singular cluster out of two bipyramids that share a vertex and have three additional distance constraints. 
Each rule gives a system of algebraic equations that is easier to solve than the complete set of distance equations. 
The rules can also be extended to floppy clusters \cite{Hauenstein:o-dSUwTH}.
Systematically investigating these ideas is a work in progress.


\section{Free energy of sticky-sphere clusters}\label{sec:thermo}

\begin{figure}
\center
\includegraphics[width=0.9\linewidth]{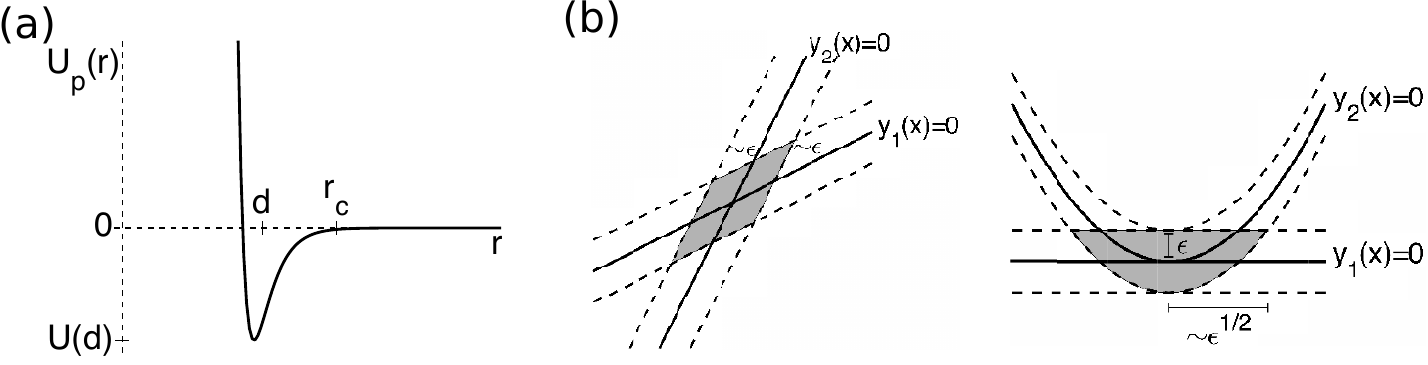}
\caption{
(A) A sketch of the pairwise potential. It has a minimum at the particle diameter $d$, decays beyond a cutoff $r>r_c$, and increases rapidly for $r<d$. 
(B) Toy models to understand the free energy of singular clusters. Left: a rigid cluster is formed at the intersection of two lines. The entropy in the sticky limit is proportional to the volume of the lines when thickened by $\eps\ll 1$, which is $\propto\eps^2$. Right: a singular cluster may be the intersection of a parabola and a tangent line. The volume of the thickened curves is $\propto\eps^{3/2}$, which goes to zero more slowly than in the regular case.
}\label{fig:sketches}
\end{figure}

\subsection{A picture of the landscape}\label{sec:landscape}

The potential energy of a sticky-sphere cluster 
is the same for all clusters with the same number of contacts, 
yet in a thermal system, the clusters they approximate can occur with vastly different frequencies. What distinguishes them is entropy -- the size of space they can explore and still keep their identity \cite{Manoharan:2015ko}. 
To calculate this entropy requires  a model for the pair potential, since a perfect delta function is not physical. Real potentials have a finite range, which makes a contribution to the entropy that does not vanish as the range decreases. 
A natural approach is to start with a particular potential and consider the Boltzmann distribution in the limit as the range goes to zero (and the depth simultaneously goes to $\infty$.)  
The limit was originally considered by Baxter \cite{Baxter:1968dh} for a square-well potential, and more recently it was considered for smoother potentials \cite{HolmesCerfon:2013jw}. 
Somewhat remarkably, the limiting entropy does not depend on the choice of potential. 

This limit also allows us to compute the entropy of floppy clusters. These clusters have internal degrees of freedom, so there is a positive-dimensional region in configuration space they can access by deforming while maintaining their contacts. On this region the potential energy is constant. 
Each region is typically a manifold, with dimension equal to the number of internal degrees of freedom of the cluster (after modding out by $SE(3)$ to obtain a quotient manifold.) 
A rigid cluster is a zero-dimensional manifold, or a point. 
If we break a bond in a rigid cluster, we obtain a cluster with one internal degree of freedom, which is a one-dimensional manifold or a line. 
Breaking two bonds gives a two-dimensional manifold, whose boundaries are the lines, and continuing up in dimension we obtain the entire energy landscape as the union of manifolds of different dimensions, glued together at their boundaries. A helpful schematic is of a high dimensional polytope, whose faces have edges, which in turn have lower-dimensional edges, and so on.\footnote{The regions are not always manifolds; in general they are algebraic varieties. The topology of the stratification is almost certainly more complicated than that of a polytope.} 
In the sticky limit, the Boltzmann distribution concentrates on each of these manifolds, becoming a sum of singular densities of different dimensions. 
Figure \ref{fig:n6}(c) shows an example of a two-dimensional manifold and its 1- and 0-dimensional boundaries.

\subsection{Partition functions in the sticky limit}\label{sec:sticky}

We describe the sticky limit for smooth potentials though the argument applies nearly verbatim for a square-well one. 
Consider a cluster with $m$ bonds as in \eqref{eq:bonds} that lives on a region  $\bar{\Omega}_{E,\iota}$ in configuration space  (the subscript  $\iota$ is included to index the  disconnected, non-isomorphic regions with the same constraints.) We assume the constraints are regular everywhere on $\bar{\Omega}_{E,\iota}$, meaning the rank of the rigidity matrix equals $3n-m$. 
We let $\Omega_{E,\iota} = \bar{\Omega}_{E,\iota}/ SE(3)$ be the quotient space formed by identifying all points that are the same up to rigid-body motions, 
and assume this quotient space is a Riemannian manifold.

We take the potential energy of a cluster to be $U(\bb{x}) = \sum_{i\neq j}U_p(|\bb{x}_i-\bb{x}_j|)$,  a sum of pair potentials $U_p(r)$ depending on distance $r$ between each pair. 
The pair potential is assumed to have a minimum at $d$, the sphere diameter, to decay rapidly to zero beyond some cutoff $r_c$, and to increase rapidly to $\infty$ for $r<d$ (Figure \ref{fig:sketches}.) 
The sticky limit occurs when the pair potential is both \emph{narrow} and  \emph{deep}. 
This can be achieved technically by shrinking the width by some parameter $\eps\ll 1$, and scaling the depth by a function $C(\eps)$, chosen so the nondimensional partition function for a single contact is constant. For finite $\eps$ this constant is proportional asymptotically to 
 \begin{equation}\label{eq:kappa}
 \kappa = \oneover{d}\frac{\sqrt{c_v}e^{-\beta U_0}}{\sqrt{\beta U''_0}},
 \end{equation} 
 where $c_v = 2\pi\: (\pi/2)$ if the potential is soft (hard), $U_0 = U_p(d)$, $U''_0 = U_p''(d)$, 
 and $\beta=(k_bT)^{-1}$ is the inverse of temperature $T$ times the Boltzmann constant. 
The constant $\kappa$ has been called the \emph{sticky parameter}, because it measures how sticky the particles are: the larger it is, the more time they like to spend in a cluster with more contacts. 
It is a natural way to measure the strength of a short-range bond: the depth by itself is misleading, because bonds break more rapidly in a narrow well. In the sticky regime, the width $w$ and Boltzmann factor 
combine to give $\kappa \approx e^{-\beta U_0}\cdot w$, which must be not too large or small for the limit to converge to a finite value. 

The partition function for $\bar{\Omega}_{E,\iota}$ is the integral of the Boltzmann distribution over a neighbourhood $N_{E,\iota}$ associated with the cluster, obtained by fattening the constraints by $\eps$ so the bonds can vibrate, allowing for translations, rotations, and possibly reflection, and including all geometrically isomorphic copies of the manifold obtained by permuting identical particles. After non-dimensionalizing lengths the partition function is  
\begin{equation}\label{eq:Zdef}
Z_{E,\iota} = \oneover{d^{3n}}\int_{\mathcal{N}_{E,\iota}} e^{-\beta U(\bb{x})}d\bb{x} .
\end{equation}
This expression is evaluated in the limit as $\eps\to 0$. 
The result, neglecting small differences in excluded volume and factors that are the same for all clusters, is 
\begin{equation}\label{eq:Zalpha}
Z_{E,\iota} = \kappa^m z_{E,\iota}\e{g},
\end{equation}
with
\begin{equation} 
z_{E,\iota,}\e{g} = \oneover{d^{3n-m}}\int_{\Omega_{E,\iota}} \frac{|\bb{I}(\bb{x})|^{1/2}}{\sigma}\prod_{i=1}^{m}\lambda_{i}^{-1/2}(\bb{x}) \mu_{E,\iota}(d\bb{x}).
\end{equation} 
The integral is with respect to the natural volume form $\mu_{E,\iota}$ on the quotient manifold \cite{HolmesCerfon:2013jw}. 
Here $\sigma$ is the symmetry number, which counts the number of permutations of identical particles that are equivalent to an overall rotation (and reflection, if entantiomers are lumped into one state.) 
The matrix $\bb{I}$ is the moment of inertia tensor formed by setting all particle masses to 1 \cite{Cates:2015ik}; the square root of its determinant is proportional to the volume of the space of rotations. 
The $\lambda_{i}$ are the non-zero eigenvalues of $R^TR$, where $R(\bb{x})$ is the rigidity matrix defined in \eqref{eq:R}. 
They arise because in the sticky limit the dynamical matrix approaches $\grad\grad U = U''_0R^TR$ and the integral over vibrational directions is evaluated in a harmonic approximation.  


The limiting partition function factors into two pieces: one is the sticky parameter, which depends on the pair potential, temperature, and particle diameter, and the other is the \emph{geometrical partition function} $z_{E,\iota}\e{g}$, so-called because it depends only on the relative positions of the spheres, but not on any system-dependent quantities. 
This separation 
has several advantages, both conceptual and practical. 
Conceptually, it is helpful because it makes transparent which parts of the partition function will change with parameters in the system, and which are fundamental properties of the particles themselves. 
For example, from the observation that clusters with the same number of bonds have the same power of $\kappa$, we see their relative probabilities must be governed purely by geometry -- they will not change with parameters such as temperature. 
Computationally, it is helpful because while calculating the integral in \eqref{eq:Zalpha} is a challenge, it only needs to be done once -- different temperatures or interaction potentials are accounted for by varying the single parameter $\kappa$. If the particles have different, specific interactions, one can easily adapt this framework by allowing the sticky parameters for different contacts to vary \cite{Perry:2016gk}. 
When some particles do not interact at all, then there are local minima that are themselves floppy \cite{Zeravcic:2014it}. 
In this case computing the integral \eqref{eq:Zalpha} is critical to understanding their entropy, since it cannot be obtained through any local approximation. 

A first step to calculating the integrals in \eqref{eq:Zalpha} was taken in Holmes-Cerfon et al \cite{HolmesCerfon:2013jw}, which calculated the 0,1,2-dimensional integrals for $n=6,7,8$ by explicitly parameterizing the manifolds (Figure \ref{fig:n6}). This is straightforward in one dimension but much less so in two. 
To calculate integrals over higher-dimensional manifolds, there are sometimes natural variables with which to parameterize such as the distances between non-bonded spheres \cite{Ozkan:2011vy}, but in general randomized methods are probably required.

\subsection{Experimental measurements of free energy}\label{sec:expt}

Experiments can isolate small collections of colloidal particles and measure the configurations they assemble into, at a level of detail completely inaccessible to atomic clusters. This has been a way to  validate the calculations above, showing they can quantitatively describe a real system, and also to point to missing ingredients, such as neglected physics or situations when the sticky limit breaks down \cite{Meng:2010gsa,Perry:2012kf,Perry:2015ku,Perry:2016gk}.   
These measurements have proven educational since colloidal systems follow the laws of classical statistical mechanics, about which there is still some confusion as they are often taught by analogy to quantum mechanics \cite{Cates:2015ik}.
Experiments have also highlighted the stark difference between sticky hard sphere and longer-range atomic clusters. 


This difference was strikingly illustrated with experiments by Meng et al \cite{Meng:2010gsa}. They isolated small numbers of 1$\mu$m colloidal spheres in microwells that interacted attractively via depletion over a range roughly 1.05 times their diameter. 
The spheres clumped up into clusters large enough to see by eye in a microscope, so Meng et al could identify the rigid cluster that each one most resembled. The observed frequency of each cluster is its equilibrium  probability, which in the sticky limit is proportional to the partition function \eqref{eq:Zalpha}. 

The experimental and theoretical occupation probabilities agreed well. The best agreement was at $n=6$ (see Figure \ref{fig:n6}), where the octahedron occurred with experimental (theoretical) frequencies 95.7\% (96\%), and the polytetrahedron with frequencies 4.3\% (4.0\%). 
This drastic difference in frequencies was itself a major discovery. 
The octahedron is more symmetric than the polytetrahedron, so would be favoured energetically in a cluster with a longer-range potential. 
Even for the Lennard-Jones 6-12 potential, often used to model short-ranged interactions, the octahedron is 0.3 units lower in energy than the polytetrahedron \cite{Hoare:1971ke}, so would be favoured at low temperature. 
In a sticky-sphere system  the frequency difference can only be attributed to entropy, which is  suppressed 
by the symmetry number. 

The theory and measurements begin to disagree for clusters with small gaps comparable to the width of the actual potential, which happens for some clusters at $n=8$.  
For $n\geq 9$ the number of samples was not large enough to obtain statistics on all clusters, but those observed point to some interesting trends. The singular cluster at $n=9$ was the most frequent by far, occurring about 10\% of the time. For $n=10$, singular and hyperstatic clusters predominated, 
with frequencies about 20\% and 10\% respectively.  
This suggests a competition between singular clusters and extra contacts as $n$ increases.

A similar set of experiments was performed by Perry et al \cite{Perry:2015ku} to analyze floppy clusters. They created a two-dimensional system in which spheres moved on a plane, like discs, and interacted through a depletion force. 
The particle locations were automatically extracted so they could identify the nearest sticky-sphere manifold and each cluster's position on it. 
Perry et al mainly studied clusters of 6 discs, for which one can verify there are three rigid clusters, all fragments of a hexagonal lattice. They measured the occupation probabilities of the floppy modes formed by breaking one and two contacts. By \eqref{eq:Zalpha}, the frequencies conditional on having a certain number of bonds broken should not depend on the potential, so can be computed despite limited knowledge of the electrostatic, van der Waals, and depletion forces that contribute. The experimentally measured frequencies agreed with those calculated from \eqref{eq:Zalpha}, showing the sticky-sphere limit applies equally to floppy clusters.  

Perry et al made another important contribution by showing that one can use these coarse-grained observations to measure $\kappa$. Typically, measuring an interaction potential, especially one that is stiff, requires high-frequency, high-resolution measurements to resolve the details of the well when two particles are nearly in contact. Predicting $\kappa$ by first measuring $U_0$, $U_0''$ would be an experimental tour de force, and estimations gave a range of $\kappa\approx 2-200$. But $\kappa$ can be inferred from the macroscopic data by observing that it governs the ratio of occupation probabilities between manifolds of different dimensions. For discs, this gives 
\begin{equation}
\frac{\text{time in rigid clusters}}{\text{time in 1-bond-broken clusters}} = \frac{\kappa^{2n-3}Z_0}{\kappa^{2n-4}Z_1} ,
\end{equation}
where
\begin{equation}
Z_i = \sum_{(E,\iota): \dim \Omega_{E,\iota} = i} z_{E,\iota}\e{g}
\end{equation}
 is the sum of the geometrical partition functions for manifolds of dimension $i$. 
 The $Z_i$ are known from the theory, and the fraction on the left-hand-side is measured experimentally, so one can solve this algebraic equation for $\kappa$. 
 By also comparing the 1 and 2-dimensional manifolds and by considering clusters of different sizes,  Perry et al found measurements in the range $\kappa\approx27-35$,  narrow enough given the measurement and statistical uncertainties. 
This method was later used to infer that particles with differing compositions had different interaction strengths, despite the interactions originating from the same depletant \cite{Perry:2016gk}.

\subsection{Free energy of singular clusters}\label{sec:singular}

The free energy of the singular cluster in Figure \ref{fig:singular} is not possible to predict using \eqref{eq:Zalpha}, because the sticky limit relies on a harmonic approximation which fails when the dynamical matrix acquires an extra zero eigenvalue. Yet, the high frequency with which this cluster was observed in experiments  suggests that degenerate vibrational degrees of freedom could contribute significantly to the entropy. 
How does this additional entropy compare with the energy of an extra contact?  

Although the sticky limit diverges for both singular and hyperstatic clusters, it may be possible to compare the two by considering the leading-order terms in an asymptotic expansion of the partition function. 
A simple example shows why. 
Suppose that ``configuration space'' is $\R^2$, and ``contacts'' are solutions to equations $y_i(x)=0$, ($i=1,2$.) 
A ``regular cluster'' is the point where two curves intersect non-tangentially, as in the solution $x=(0,0)$ to  $y_i(x)  = v_i\cdot x =0$ where $v_1,v_2\in \R^2$ are linearly independent (Figure \ref{fig:sketches}). For a square-well potential with width $\eps$ and depth $U_0$ the partition function is the integral of the Boltzmann factor over the region $\Omega=\{x: |y_1(x)|,|y_2(x)| < \eps\}$, which equals $4e^{-\beta U_0}|v_1\times v_2|^{-1} \eps^2$. This is $O(\eps^2)$ as $\eps \to 0$, as expected since the volume is two-dimensional. 

A ``singular cluster'' is formed when curves intersect tangentially, such as the  intersection $x=(0,0)$ of a line $y_1(x) =x_2=0$ and a parabola $y_2(x) = x_1^2-x_2=0$. 
 The integral of the Boltzmann factor over a region of the form $\Omega$ can be shown to be $O(\eps^{3/2})$ as $\eps\to0$:  it goes to zero \emph{more slowly} than that for a regular cluster. 
In the sticky limit $U_0$ is scaled so the partition function for a regular cluster approaches an $O(1)$ constant, so the partition function for a singular cluster will blow up. 
However, this toy calculation shows that the leading-order contribution to the partition function is entirely computable and should depend on both the sticky parameter and one more parameter characterizing the width of the potential. 
Calculations extending this argument to clusters that are second-order rigid have since been published in \cite{Kallus:bQpIXJhw}. 

\section{Kinetics}\label{sec:kinetics}

\begin{figure}
\center
\includegraphics[width=\linewidth]{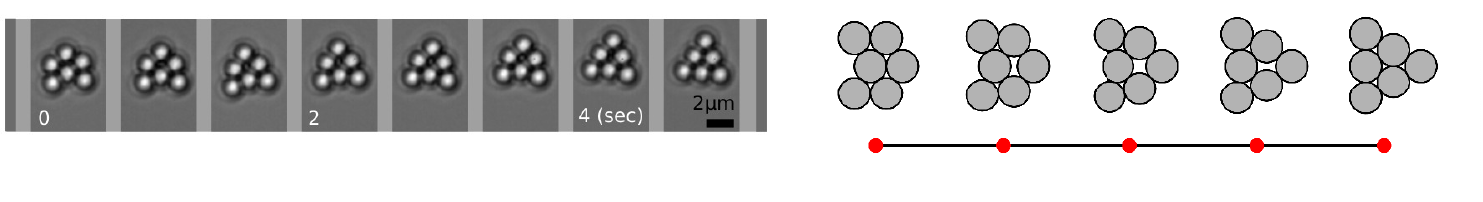}
\caption{Transitions happen diffusively along one-dimensional paths. (A) A transition observed  experimentally between rigid clusters of discs. (B) Sketch of selected states on the transition path. The line segment representing the path is shown with corresponding points in red. Part (A) adapted with permission from \cite{Perry:2015ku}. Copyrighted by the American Physical Society.}\label{fig:transition}
\end{figure}

When a colloidal cluster in a rigid state breaks a bond, it doesn't immediately form another one. Rather, it wiggles and jiggles its way around its floppy degree of freedom, sometimes coming close to the original rigid state, sometimes coming close to a different one, before eventually falling into a well when two particles come into contact (Figure \ref{fig:transition}).  Describing this process and the rate with which it occurs requires more than simply the height of the energy barrier for breaking a bond; we need to understand the diffusive process in between. 

The sticky limit gives a way to do this. If we model a system with the overdamped Langevin dynamics, then we can apply the limit of a deep, narrow potential 
 to the Fokker-Planck equation describing the evolution of the probability density. The limiting equation is a system of coupled Fokker-Planck equations, one on each manifold that forms the energy landscape, describing the flow of probability along each manifold and the flux to others in and out of their boundaries. This system is a complete description of the dynamics in the sticky limit, and provides a natural starting point to describe dynamic phenomena like transition rates between ground states, assembly pathways to reach the ground states, epitaxy, defect motion, nucleation, growth, among many others.

\subsection{Theoretical calculations}

The sticky limit of the Fokker-Planck equation is computed using an asymptotic procedure akin to boundary layer theory \cite{HolmesCerfon:2013jw}. 
Assuming constant, diagonal diffusivity $D$, the limiting equation on manifold $\Omega_{E,\iota}$ with $m$ contacts is 
\begin{multline}\label{eq:FPcluster}
\partial_t P_{E,\iota} = \\
D\text{ div}_{E,\iota} \left(-P_{E,\iota}\underbrace{\text{ grad}_{E,\iota} \log h_{E,\iota}}_{\text{effective force}} + \underbrace{\text{ grad}_{E,\iota}P_{E,\iota}}_{\text{diffusion}} \right)  + \underbrace{\kappa^{-1} \sum_{(F,\nu)\to(E,\iota)} j_{F,\nu}\cdot \hat{n} }_{\text{flux to/from $\Omega_{F,\nu}$}}.
\end{multline}
Here $h_{E,\iota}(\bb{x}) = |\bb{I}(\bb{x})|^{1/2}\prod_{i=1}^{m}\lambda_{i}^{-1/2}(\bb{x})$ is the integrand in \eqref{eq:Zalpha}, 
and $P_{E,\iota}(\bb{x},t) = h_{E,\iota}(\bb{x}) p(\bb{x},t)$ is a density (with respect to the natural quotient volume form) on manifold $\Omega_{E,\iota}$, 
from which the actual probability density on $\Omega_{E,\iota}$  is calculated as $\kappa^{m}P_{E,\iota}$. Function $p(\bb{x},t)$ is defined everywhere and is the density of the probability distribution 
with respect to the equilibrium probability distribution. 
The operators $\text{grad}_{E,\iota}$, $\text{div}_{E,\iota}$ are the gradient and divergence with respect to the natural quotient metric on each manifold. 
The final term is a sum over fluxes $j_{F,\nu} = -D\left(-P_{F,\nu} \text{grad}_{F,\nu} \log h_{F,\nu} + \text{ grad}_{F,\nu}P_{F,\nu}\right)$ such that  $\Omega_{E,\iota}$ is part of the boundary of $\Omega_{F,\nu}$, with $\text{dim}(\Omega_{F,\nu}) = \text{dim}(\Omega_{E,\iota}) +1$, and $\hat{n}$ is an outward normal vector. 
System \eqref{eq:FPcluster} does not yet lump together geometrically isomorphic manifolds so the index $\iota$ now includes all copies of the manifold obtained by permuting particles. 

We call \eqref{eq:FPcluster} the ``sticky Fokker-Planck equations'', because they describe a generalization of a sticky Brownian motion, which is a Brownian motion that has been slowed down on a boundary in such a way that it spends a non-zero amount of time there \cite{Ikeda:1981}. In the simplest case where a particle diffuses on the half-line $[0,\infty)$ with a sticky point at the origin, the sticky Fokker-Planck equations would be $p_t = p_{xx}$ with boundary condition $\kappa p_t(0) =  p_x(0)$ or equivalently $\kappa p_{xx}(0) = p_x(0)$. Similarly substituting for time derivatives in \eqref{eq:FPcluster} shows it is really a hierarchy of second-order boundary conditions.

The probability in the interior of each manifold dynamically evolves due to three terms: diffusion on the manifold, and forcing on the manifold, and flux from higher-dimensional manifolds. 
The forcing is entropic  and arises because the vibrational and rotational entropies change along the manifold; it is the same force obtained by considering a harmonic potential constraining the system near the manifold \cite{Ciccotti:2007fv}.

\subsection{Transition Rates}

\begin{figure}
\center
\includegraphics[width=0.9\linewidth]{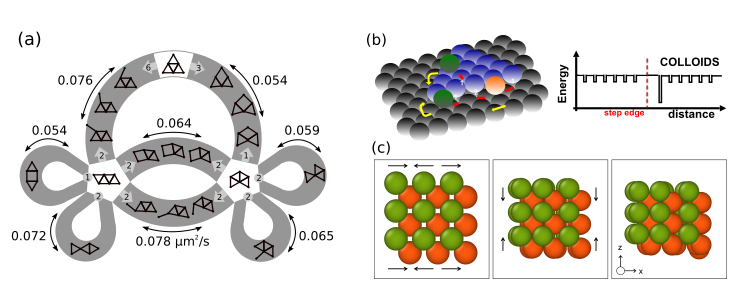}
\caption{ Kinetics in the sticky limit are strongly influenced by diffusion. (A) Diffusion coefficients measured for each possible transition path between rigid clusters of 6 discs. (B) Schematic of colloidal epitaxy, showing a colloid diffusing between sites on a lattice. 
The free energy barrier is strongly influenced by the length of the path, which is longest at a step edge. 
(C) A bcc crystallite of two kinds of particles (viewed along (010) axis), transforming along a diffusive pathway to an fcc fragment.  
(A) adapted with permission from \cite{Perry:2015ku}. Copyrighted by the American Physical Society. 
(B) adapted from \cite{Ganapathy:2010be}. Reprinted with permission from AAAS. 
(C)  adapted with permission from \cite{Jenkins:2014js}. 
}\label{fig:kinetics}
\end{figure}

If the sticky parameter $\kappa$ is large, then we expect a cluster to spend most of its time in equilibrium as a rigid cluster, only occasionally changing shape to another cluster. How and how often do transitions occur? 
Intuitively, we might expect a transition happens by a cluster breaking a single bond, and diffusing along its one-dimensional degree of freedom until it forms another bond at the other end. If so, the rate should be determined by the timescale to diffuse along a line segment. Indeed, this idea was used 
by Perry et al \cite{Perry:2012kf}, without reference to a sticky limit, to estimate the timescale to transition between an octahedron and polytetrahedron, and earlier by Ganapathy et al \cite{Ganapathy:2010be}  to describe a colloid hoping between sites on a hexagonal lattice during epitaxy. 

\subsubsection{Theoretical transition rates} The picture above is asymptotically correct when the sticky parameter is large. 
Transition rates can be calculated exactly from a solution to the backward Fokker-Planck equation using Transition Path Theory \cite{E:2010hs,VandenEijnden:2014ita}. 
Solving directly is hopeless because the equations \eqref{eq:FPcluster} are all coupled, from the lowest to the highest dimensions, but when $\kappa$ is large they separate asymptotically. To leading order in $\kappa^{-1}$ the frequency of transition between rigid clusters $A$ and $B$ is obtained from the flux of probability along the one-dimensional paths that connect them, as \cite{HolmesCerfon:2013jw}
\begin{equation}\label{eq:kab} 
\nu_{AB} = \kappa^{-1}\frac{D}{d^2}Z_0^{-1}\sum_{(E,\iota)} Q_{E,\iota}^{-1}, \qquad Q_{E,\iota} = \int_{\Omega_{E,\iota}} h_{E,\iota}^{-1} ds,
\end{equation} 
where the sum is over all $(E,\iota)$ such that $\Omega_{E,\iota}$ is a one-dimensional manifold connecting cluster $A$ to cluster $B$, and $s$ is an arc-length parameterization of $\Omega_{E,\iota}$. 
This frequency is the average number of times a transition between $A$ and $B$ will be observed in equilibrium, and is related to the rate of leaving a certain state to leading order as $k_{AB} = \nu_{AB}/(z_A\e{g}/Z_0)$ \cite{VandenEijnden:2014ita}.

This expression again conveniently separates into a geometrical part, that can be pre-computed, and a set of constants that depend on parameters in the system. 
It is expected to be more accurate than rates computed from properties of saddle points, as in Transition State Theory \cite{Wales:2003,VandenEijnden:2005fs}. These predict rates of the form $k^s_{AB} \propto \frac{\beta^{-1}z_s}{z_A\e{g}/Z}e^{- \beta U_0}$, where $z_s$ is a pre-factor depending on properties of the saddle point, such as its vibrational partition function, and $Z$ is the total partition function. 
While the Arrhenius factors are the same, 
there is no reason why $z_s$ should bear much relation to the geometric factors in \eqref{eq:kab}, something which has been confirmed through numerical tests by the author and collaborators.

\subsubsection{Experimental measurements of dynamics} 
These computations were tested directly in experiments by Perry et al \cite{Perry:2015ku}, which counted the number of transitions observed between each pair of rigid clusters of six discs on a plane. These numbers can be directly compared to the theoretical prediction \eqref{eq:kab} by substituting the values of the constants. 
The particle diameter $d$ is known in advance, and the sticky parameter $\kappa$ was measured separately in section \ref{sec:expt}, but the 
the particle diffusivity $D$ posed a problem. Substituting values for the single particle diffusivity in an unbounded three-dimensional domain, as well as near a two-dimensional wall, gave predicted rates that were roughly 6 and 2 times too big, respectively. 
This is because the collective motion of the discs during a transition creates a hydrodynamic flow that alters the discs' mobility, and hence, by Batchelor's generalization of the Stokes-Einstein relation, their diffusivity tensor \cite{Batchelor:1976gz,Dufresne:2000bg}. 
Fortunately, for such a low-dimensional motion the average component of the diffusivity tensor along each transition path can be measured from the time series of each transition. 
Substituting either the measured diffusivity or incorporating the different measured values for each path gave results that agreed with the measured transition rates. 
This shows that the limiting sticky dynamics can predict experimental transition rates, but that accounting for hydrodynamic interactions is critical for obtaining quantitative agreement. 

The importance of hydrodynamics in sticky-particle assemblies was also highlighted by Jenkins et al \cite{Jenkins:2014js}, to explain the transition observed in a crystal of DNA-linked particles from a bcc phase to an fcc phase upon annealing. The puzzle is that the energy of all close-packings is the same, and entropy overwhelmingly favours random stackings of hexagonal planes. So why should the entropically unlikely fcc phase be the first one that is formed?  Jenkins et al argued that if one thinks of particles as sticky, then the bcc phase is a floppy manifold with a great many degrees of freedom. Most of these lead nearly nowhere since particles collide, but some degrees of freedom -- those that are a special combination of  sliding planes -- can be extended much farther. 
Jenkins et al likened this manifold to a bicycle wheel, with a small fat hub near the bcc phase, and several long thin spokes leading out.  
To explain why the system chooses the rare spokes that lead to an fcc phase, Jenkins et al computed the hydrodynamic mobility along a representative sample of spokes and showed that it was more than 50 times higher for the those leading to the fcc phase than for those leading to random stackings. 
They argued that although fcc is not the most thermodynamically stable, it is the most kinetically accessible so is the one seen on the timescales of the experiment. 

%



\section{Outlook}\label{sec:outlook}

The sticky limit predicts states, free energies and transition rates of clusters that agree with those observed for colloidal clusters, and 
the hope is that it will give insight into a wider range of phenomena, both in clusters and also in bulk systems like crystals made of DNA-coated particles. 
For this to happen requires developing not only computational tools 
to work with different particle shapes, sizes, and interaction structures, and  larger or higher-dimensional systems, 
but also developing theoretical tools. 
New approximations are required to describe a wider variety of dynamic phenomena, like nucleation,  assembly pathways,  and out-of-equilibrium growth processes (e.g. \cite{Chen:2011bj}).
The sticky Fokker-Planck equations are a starting point, but are too high-dimensional to work with directly. The approximations are expected to be different from those used in traditional energy landscape theories, and if they can maintain a separation between input parameters and geometry, may lead to efficient methods to solve inverse problems such as designing a system that self-assembles both reliably and efficiently. 

Another issue is to extend the sticky limit and surrounding computational apparatus to singular clusters, which are persistent features of the landscape for $n\geq 9$ spheres.  
Only when these are incorporated will it be possible to address the question of emergence, and determine how close-packings come to dominate the landscape for large $n$ despite being disfavoured by symmetry.
Even if singular clusters do not end up being the most thermodynamically stable states, they could play a role in kinetic effects like transitions, or lead to interesting bifurcations as the geometrical parameters in the system are changed. 

For this theory to make specific, testable predictions also requires incorporating the relevant physics. Hydrodynamic interactions are critical in determining the kinetics but are difficult to measure except on low-dimensional paths. Sticky tethers like DNA could also influence the kinetics, but exactly how is not well understood 
\cite{Xu:2011gs,Mani:2012dia,Rogers:2013dc}, 
and nor is the impact of surface friction, such as created by particle roughness \cite{Still:2014bd,HolmesCerfon:2016wu}. Discrepancies from the sticky-limit predictions can help identify missing physics, but models are needed to make predictions for larger systems that can't be directly measured.

The set of rigid clusters by itself has already proven useful in studying phenomena like self-assembly and self-replication. Because it is a nearly complete set of local minima on a particular landscape, it is a toy model that realistically captures the geometrical frustration experienced by physical and biological systems. It  
 has been used to ask questions like: how does one make a particular rigid cluster the most thermodynamically stable, if all one can change are the interaction strengths and specificities \cite{Tkachenko:2011iv,Hormoz:2011ir,Zeravcic:2014it,Miskin:2016bj}? 
How can one make a cluster that reproduces itself \cite{Zeravcic:2014ev}? 
It is natural to work in the sticky limit, because one is interested in comparing interaction strengths and structures, but not in the detailed shape of the energy landscape. 
The computational apparatus surrounding the sticky limit is expected to 
provide a concrete tool to make forward or inverse predictions incorporating specific experimental constraints.



The idea of particles bound by distance constraints that are possibly harmonic has been used to study  a number of others condensed-matter systems such as jamming \cite{OHern:2002bsa}, structural glasses \cite{Boolchand:2005bi}, and silicates \cite{Hammonds:1996wy}. 
These systems have singularities, like clusters, which have been evoked to explain behaviour near critical points 
\cite{Wyart:2007dh,Xu:2010fa,Gomez:2012wc}. 
The properties of these systems as frameworks govern many of their bulk behaviours, so a new thrust in materials science has been to solve the inverse problem, of designing a framework that responds in a desired way to stress. This might be possible by engineering it to have soft modes with localized spatial deformations or other, possibly nonlinear, properties \cite{Kane:2013if,Chen:2014ec}.  
So far the procedure has been to design modes by hand, such as by twisting units in a kagome lattice \cite{Paulose:2015dc}. However, if the set of possible frameworks can be automatically enumerated, as they can for clusters, this opens the door to a richer set of materials. 
New materials or structures may also be assembled like origami, by patterning a two-dimensional surface so it can bend and fold  \cite{Pandey:2011jj,Silverberg:2014dn,Sussman:2015ex}. 
Each facet of the surface is an object that is bound to the others by distance and angle constraints, so its configuration space resembles that of clusters \cite{Demaine:2007jh}. As these material systems become smaller, thermal effects will become important, and the tools developed for clusters may be useful.

Of course, the sticky limit never holds exactly since a real potential has a finite range, and this leads to discrepancies between predicted and measured free energies even for clusters as small as $n=8$. An exciting possibility is whether the sticky limit can be used as a starting point to understand the landscapes of finite-range potentials. 
One can imagine starting with the sticky-sphere landscape,  slowly turning on a givenpotential, and 
relaxing the landscape in some manner. 
The hope is we could find all the pieces of the final landscape, and more efficiently than exploring it from scratch; indeed the landscape for a short-ranged potential is thought to be the most rugged, with fewer local minima as the range increases \cite{Hoare:1976bb,Miller:1999ct,Wales:2001cq}. 
Such a continuation would give insight into why a landscape has a particular shape, and may also provide a bound on the space of possible landscapes; for example, the space of energy-minimizing configurations of points on the sphere is sometimes much lower-dimensional than the space of interaction potentials \cite{Ballinger:2009jj}.
These and the ideas above may make the sticky limit a powerful starting point for understanding more general energy landscapes.

\section*{ACKNOWLEDGMENTS}
I would like to thank Michael Brenner, Louis Theran, Steven Gortler,  Yoav Kallus,  John Crocker,  Vinothan Manoharan, and Eric Vanden-Eijnden for helpful discussions, and Michael Brenner and Bill Holmes for detailed comments on previous drafts. 
Research discussed from my own group is funded through grant DE-SC0012296. 
Posted with permission from the Annual Review of Condensed Matter Physics, Volume 8, 2017 by Annual Reviews,  
http://www.annualreviews.org.



\begin{thebibliography}{97}

\bibitem{Frank:1952eg}
Frank FC. 1952 215:43--46

\bibitem{Stillinger:1984tr}
Stillinger FH, Weber TA. 1984.
\textit{Science} 225:983--989

\bibitem{Nelson:1989ij}
Nelson DR, Spaepen F. 1989.
\textit{Solid State Physics} 42:1--90

\bibitem{Doye:1996vn}
Doye J, Wales DJ. 1996.
\textit{Science} 271:484--487

\bibitem{Sedgwick:2004di}
Sedgwick H, Egelhaaf SU, Poon WCK. 2004.
\textit{J. Phys.: Condens. Matter} 16:S4913--S4922

\bibitem{PatrickRoyall:2008fz}
Patrick~Royall C, Williams SR, Ohtsuka T, Tanaka H. 2008.
\textit{Nat Mater} 7:556--561

\bibitem{Yunker:2009jd}
Yunker P, Zhang Z, Aptowicz KB, Yodh AG. 2009.
\textit{Phys. Rev. Lett.} 103:115701

\bibitem{Ganapathy:2010be}
Ganapathy R, Buckley MR, Gerbode SJ, Cohen I. 2010.
\textit{Science} 327:445--448

\bibitem{Wang:2011gv}
Wang T, Sha R, Dreyfus R, Leunissen ME, Maass C, et~al. 2011.
\textit{Nature} 478:225--228

\bibitem{Zeravcic:2014ev}
Zeravcic Z, Brenner MP. 2014.
\textit{Proc. Natl. Acad. Sci.} 111:1748--1753

\bibitem{Manoharan:2004jk}
Manoharan VN, Pine DJ. 2004.
\textit{MRS Bull} 29:91--95

\bibitem{Fan:2010jn}
Fan JA, Wu C, Bao K, Bao J, Bardhan R, et~al. 2010.
\textit{Science} 328:1135--1138

\bibitem{Hormoz:2011ir}
Hormoz S, Brenner MP. 2011.
\textit{Proc. Natl. Acad. Sci.} 108:5193--5198

\bibitem{Schade:2013ee}
Schade N, Holmes-Cerfon M, Chen E, Aronzon D, Collins J, et~al. 2013.
\textit{Phys. Rev. Lett.} 110:148303

\bibitem{Zeravcic:2014it}
Zeravcic Z, Manoharan VN, Brenner MP. 2014.
\textit{Proc. Natl. Acad. Sci.} 111:15918--15923

\bibitem{Lu:2013dn}
Lu~陸述義 PJ, Weitz DA. 2013.
\textit{Annu. Rev. Condens. Matter Phys.} 4:217--233

\bibitem{Manoharan:2015ko}
Manoharan VN. 2015.
\textit{Science} 349:1253751--1253751

\bibitem{Asakura:1954jy}
Asakura S, Oosawa F. 1954.
\textit{J. Chem. Phys.} 22:1255--1256

\bibitem{Nykypanchuk:2008cp}
Nykypanchuk D, Maye MM, van~der Lelie D, Gang O. 2008.
\textit{Nature} 451:549--552

\bibitem{Dreyfus:2009gl}
Dreyfus R, Leunissen M, Sha R, Tkachenko A, Seeman N, et~al. 2009.
\textit{Phys. Rev. Lett.} 102:048301

\bibitem{Rogers:2011et}
Rogers WB, Crocker JC. 2011.
\textit{Proc. Natl. Acad. Sci.} 108:15687--15692

\bibitem{Macfarlane:2011fh}
Macfarlane RJ, Lee B, Jones MR, Harris N, Schatz GC, Mirkin CA. 2011.
\textit{Science} 334:204--208

\bibitem{DiMichele:2013bw}
Di~Michele L, Eiser E. 2013.
\textit{Phys. Chem. Chem. Phys.} 15:3115--3129

\bibitem{Sacanna:2013ge}
Sacanna S, Pine DJ, Yi GR. 2013.
\textit{Soft Matter} 9:8096

\bibitem{MartinezVeracoechea:2011eoa}
Martinez-Veracoechea FJ, Mladek BM, Tkachenko AV, Frenkel D. 2011.
\textit{Phys. Rev. Lett.} 107:045902

\bibitem{Romano:2012bz}
Romano F, Sciortino F. 2012.
\textit{Nat Comms} 3:975

\bibitem{Li:2013he}
Li Ting, Sknepnek R, Olvera de~la Cruz M. 2013.
\textit{J. Am. Chem. Soc.} 135:8535--8541

\bibitem{Millan:2014de}
Millan JA, Ortiz D, van Anders G, Glotzer SC. 2014.
\textit{ACS Nano} 8:2918--2928

\bibitem{OHern:2002bsa}
O{\textquoteright}Hern CS, Langer SA, Liu AJ, Nagel SR. 2002.
\textit{Phys. Rev. Lett.} 88:075507--4

\bibitem{Boolchand:2005bi}
Boolchand P, Lucovsky G, Phillips JC, Thorpe MF. 2005.
\textit{Philosophical Magazine} 85:3823--3838

\bibitem{Hammonds:1996wy}
Hammonds KD, Dove MT, Giddy AP, Heine V, Winkler B. 1996.
\textit{American Mineralogist} 81:1057--1079

\bibitem{Pandey:2011jj}
Pandey S, Ewing M, Kunas A, Nguyen N, Gracias DH, Menon G. 2011.
\textit{Proc. Natl. Acad. Sci.} 108:19885--19890

\bibitem{Demaine:2007jh}
Demaine ED, O'Rourke J. 2007.
{Geometric folding algorithms}.
Linkages, Origami, Polyhedra. Cambridge: Cambridge University Press, Cambridge

\bibitem{Silverberg:2014dn}
Silverberg JL, Evans AA, McLeod L, Hayward RC, Hull T, et~al. 2014.
\textit{Science} 345:647--650

\bibitem{Wales:2012dd}
Wales DJ. 2012.
\textit{Philosophical Transactions of the Royal Society A: Mathematical,
  Physical and Engineering Sciences} 370:2877--2899

\bibitem{Wales:2003}
Wales D. 2003.
{Energy Landscapes}.
Applications to Clusters, Biomolecules and Glasses. Cambridge University Press

\bibitem{HolmesCerfon:2013jw}
Holmes-Cerfon M, Gortler SJ, Brenner MP. 2013.
\textit{Proc. Natl. Acad. Sci.} 110:E5--E14

\bibitem{Arkus:2009dc}
Arkus N, Manoharan V, Brenner M. 2009.
\textit{Phys. Rev. Lett.} 103:118303

\bibitem{Bezdek:2012if}
Bezdek K. 2012.
\textit{Discrete Comput Geom} 48:298--309

\bibitem{Bezdek:2013de}
Bezdek K, Reid S. 2013.
\textit{J. Geom.} 104:57--83

\bibitem{Asimow:1978en}
Asimow L, Roth B. 1978.
\textit{Trans. Amer. Math. Soc.} 245:279--289

\bibitem{Connelly:1996vj}
Connelly R, Whiteley W. 1996.
\textit{SIAM Journal on Discrete Mathematics} 9:453--491

\bibitem{Connelly:2015dp}
Connelly R, Gortler SJ. 2015.
\textit{Discrete Comput Geom} 53:847--877

\bibitem{Lubensky:2015jd}
Lubensky TC, Kane CL, Mao X, Souslov A, Sun K. 2015.
\textit{Reports on Progress in Physics} 78

\bibitem{Borcea:2004dr}
Borcea C, Streinu I. 2004.
\textit{Discrete Comput Geom} 31:287--303

\bibitem{Calladine:1978dy}
Calladine CR. 1978.
\textit{International Journal of Solids and Structures} 14:161--172

\bibitem{Maxwell:1864vl}
Maxwell JC. 1864 27:294--299

\bibitem{Meng:2010gsa}
Meng G, Arkus N, Brenner MP, Manoharan VN. 2010.
\textit{Science} 327:560--563

\bibitem{HolmesCerfon:2016wa}
Holmes-Cerfon MC. 2016a.
\textit{SIAM Rev.} 58:229--244

\bibitem{Echt:1981bw}
Echt O, Sattler K, Recknagel E. 1981.
\textit{Phys. Rev. Lett.} 47:1121--1124

\bibitem{Raoult:1989dl}
Raoult B, Farges J, De~Feraudy MF, Torchet G. 1989.
\textit{Philosophical Magazine Part B} 60:881--906

\bibitem{Wales:1997jj}
Wales DJ, Doye J. 1997.
\textit{J. Phys. Chem. A} 101:5111--5116

\bibitem{Wales:2010jp}
Wales DJ. 2010.
\textit{Chem. Eur. J. of Chem. Phys.} 11:2491--2494

\bibitem{Malins:2009dt}
Malins A, Williams SR, Eggers J, Tanaka H, Royall CP. 2009.
\textit{J. Phys.: Condens. Matter} 21:425103

\bibitem{Calvo:2012bw}
Calvo F, Doye JPK, Wales DJ. 2012.
\textit{Nanoscale} 4:1085

\bibitem{Morgan:2014fw}
Morgan JWR, Wales DJ. 2014.
\textit{Nanoscale} 6:10717--10726

\bibitem{Arkus:2011tl}
Arkus N, Manoharan VN, Brenner MP. 2011.
\textit{SIAM Journal on Discrete Mathematics} 25:1860--1901

\bibitem{Hoy:2012cr}
Hoy RS, Harwayne-Gidansky J, O{\textquoteright}Hern CS. 2012.
\textit{Phys. Rev. E} 85:051403

\bibitem{Hayes:2012ty}
Hayes B. 2012.
\textit{Am Sci} 100:442--449

\bibitem{Hoy:2015hz}
Hoy RS. 2015.
\textit{Phys. Rev. E} 91:012303--7

\bibitem{Wampler:HRGv7QI6}
Wampler CW. 2012.
\textit{Handwritten notes}

\bibitem{Hoare:1976bb}
Hoare MR, McInnes J. 1976.
\textit{Faraday Discuss. Chem. Soc.}

\bibitem{Hauenstein:o-dSUwTH}
Hauenstein JD, Wampler CW. 2015

\bibitem{Baxter:1968dh}
Baxter RJ. 1968.
\textit{J. Chem. Phys.} 49:2770

\bibitem{Cates:2015ik}
Cates ME, Manoharan VN. 2015.
\textit{Soft Matter} 11:6538--6546

\bibitem{Perry:2016gk}
Perry RW, Manoharan VN. 2016.
\textit{Soft Matter} 12:2868--2876

\bibitem{Ozkan:2011vy}
Ozkan A, Sitharam M. 2011.
\textit{BICoB}

\bibitem{Perry:2012kf}
Perry RW, Meng G, Dimiduk TG, Fung J, Manoharan VN. 2012.
\textit{Faraday Discuss.} 159:211--234

\bibitem{Perry:2015ku}
Perry RW, Holmes-Cerfon MC, Brenner MP, Manoharan VN. 2015.
\textit{Phys. Rev. Lett.} 114:228301--5

\bibitem{Hoare:1971ke}
Hoare MR, Pal P. 1971.
\textit{Advances in Physics} 20:161--196

\bibitem{Kallus:bQpIXJhw}
Kallus Y, Holmes-Cerfon MC. 2016.
\textit{Phys. Rev. E}  95:022130

\bibitem{Ikeda:1981}
Ikeda N, Watanabe S. 1981.
{Stochastic Differential Equations and Diffusion Processes}.
Elsevier

\bibitem{Ciccotti:2007fv}
Ciccotti G, Leli{\`e}vre T, Vanden-Eijnden E. 2007.
\textit{Communications on Pure and Applied Mathematics} 61:371--408

\bibitem{Jenkins:2014js}
Jenkins IC, Casey MT, McGinley JT, Crocker JC, Sinno T. 2014.
\textit{Proc. Natl. Acad. Sci.} 111:4803--4808

\bibitem{E:2010hs}
E W, Vanden-Eijnden E. 2010.
\textit{Annu. Rev. Phys. Chem.} 61:391--420

\bibitem{VandenEijnden:2014ita}
Vanden-Eijnden E. 2014.
{Transition Path Theory}. In \textit{An Introduction to Markov State Models and
  Their Application to Long Timescale Molecular Simulation}. Dordrecht:
  Springer Netherlands,  91--100

\bibitem{VandenEijnden:2005fs}
Vanden-Eijnden E. 2005.
\textit{J. Chem. Phys.} 123:184103

\bibitem{Batchelor:1976gz}
Batchelor GK. 1976.
\textit{J. Fluid Mech.} 74:1--29

\bibitem{Dufresne:2000bg}
Dufresne ER, Squires TM, Brenner MP, Grier DG. 2000.
\textit{Phys. Rev. Lett.} 85:3317--3320

\bibitem{Chen:2011bj}
Chen Q, Whitmer JK, Jiang S, Bae SC, Luijten E, Granick S. 2011.
\textit{Science} 331:199--202

\bibitem{Xu:2011gs}
Xu Q, Feng L, Sha R, Seeman NC, Chaikin PM. 2011.
\textit{Phys. Rev. Lett.} 106

\bibitem{Mani:2012dia}
Mani M, Gopinath A, Mahadevan L. 2012.
\textit{Phys. Rev. Lett.} 108:226104--14

\bibitem{Rogers:2013dc}
Rogers WB, Sinno T, Crocker JC. 2013.
\textit{Soft Matter} 9:6412--6417

\bibitem{Still:2014bd}
Still T, Goodrich CP, Chen K, Yunker PJ, Schoenholz S, et~al. 2014.
\textit{Phys. Rev. E} 89:012301--5

\bibitem{HolmesCerfon:2016wu}
Holmes-Cerfon MC. 2016b.
\textit{Phys. Rev. E}  94:052112

\bibitem{Tkachenko:2011iv}
Tkachenko AV. 2011.
\textit{Phys. Rev. Lett.} 106:255501

\bibitem{Miskin:2016bj}
Miskin MZ, Khaira G, de~Pablo JJ, Jaeger HM. 2016.
\textit{Proc. Natl. Acad. Sci.} 113:34--39

\bibitem{Wyart:2007dh}
Wyart M, Nagel SR, Witten TA. 2007.
\textit{EPL} 72:486--492

\bibitem{Xu:2010fa}
Xu N, Vitelli V, Liu AJ, Nagel SR. 2010.
\textit{EPL} 90:56001--7

\bibitem{Gomez:2012wc}
Gomez LR, Turner AM, van Hecke M, Vitelli V. 2012.
\textit{Phys. Rev. Lett.} 108:058001

\bibitem{Kane:2013if}
Kane CL, Lubensky TC. 2013.
\textit{Nat Phys} 10:39--45

\bibitem{Chen:2014ec}
Chen BGg, Upadhyaya N, Vitelli V. 2014.
\textit{Proc. Natl. Acad. Sci.} 111:13004--13009

\bibitem{Paulose:2015dc}
Paulose J, Chen BGg, Vitelli V. 2015.
\textit{Nat Phys} 11:153--156

\bibitem{Sussman:2015ex}
Sussman DM, Cho Y, Castle T, Gong X, Jung E, et~al. 2015.
\textit{Proc. Natl. Acad. Sci.} 112:7449--7453

\bibitem{Miller:1999ct}
Miller MA, Doye JPK, Wales DJ. 1999.
\textit{J. Chem. Phys.} 110:328

\bibitem{Wales:2001cq}
Wales DJ. 2001.
\textit{Science} 293:2067--2070

\bibitem{Ballinger:2009jj}
Ballinger B, Blekherman G, Cohn H, Giansiracusa N, Kelly E, Sch{\"u}rmann A.
  2009.
\textit{Experiment. Math.} 18:257--283

\end{thebibliography}

\bibliographystyle{spmpsci}

\end{document}